\documentclass[showpacs,twocolumn]{revtex4-1}

\usepackage[english]{babel}
\usepackage[utf8]{inputenc}
\usepackage{graphicx}
\usepackage{epstopdf}
\usepackage{amsmath, amssymb}
\usepackage{dcolumn}
\usepackage{subfigure}
\usepackage{braket}

\DeclareMathOperator{\Tr}{Tr}

\begin{document}

\title{\bf Alice and Bob in an anisotropic expanding spacetime}

\author{Helder A. S. Costa }\email[]{hascosta@ufpi.edu.br}

\author{Paulo R. S. Carvalho }\email[prscarvalho@ufpi.edu.br]{ }

\affiliation{  Universidade Federal do Piau\'{\i}, Departamento de F\'{i}sica, 64049-550, Teresina, PI, Brazil}

\begin{abstract}

 We investigate a quantum teleportation process between two comoving observers Alice and Bob in an anisotropic expanding spacetime. In this model, we calculate the fidelity of teleportation and we noted an oscillation of its spectrum as a function of the azimuthal angle. We found that for the polar angle $\phi = \frac{\pi}{2}$ and the azimuthal angle $\theta \neq \frac{3\pi}{4} + n\pi$ with $n = 0, 1, 2, ...$ the efficiency of the process decreases, i.e., the fidelity is less than one. In addition, it is shown that the anisotropic effects on the fidelity becomes more significative in the regime of smooth expansion and the limit of massless particles. On the other hand, the influence of curvature coupling becomes noticeable in the regime of fast expansion (values of $\frac{\rho}{\omega} \gg 1$).

\end{abstract}

\maketitle

\section{Introduction}

 Quantum teleportation is a fundamental tool for the transmission of quantum information \cite{Bennett, Bouwmeester} where a sender (Alice) transmits an unknown quantum state to a remote receiver (Bob) via a quantum channel. In a perfect scheme, the sharing of a maximally entangled states between the sender and the receiver is a good example of quantum channel \cite{Horodecki}. However, in \cite{Alsing01, Bruschi01, Bruschi02} was showed that the entanglement is an observer dependent quantity and sensible to noinertial motion, which can result in loss of coherence.  
 
  Recently,  motivated by the development of relativistic quantum information \cite{Peres}, a series of theoretical studies has showed that the standard teleportation protocol of Bennett $et$ $al.$ (also called the BBCJPW protocol) \cite{Bennett} present peculiar behave in a relativistic setting \cite{Alsing02, Matsas, Jun Feng, Ge}. For example, in \cite{Alsing02, Matsas} the quantum teleportation between two observers in relativistic motion (Alice at rest and Bob uniformly accelerated) was investigate in order to study how fidelity is considerably reduced by the Unruh effect \cite{Unruh}. Moreover, in \cite{Ge} was showed that the fidelity of a teleported state between Alice who is far from the horizon and Bob locates near the event horizon of a Schwarzschild black hole is degraded because of the Hawking effect \cite{Hawking}. It is worth mentioning that the relevance of these investigation has been to understand some quantum gravitational effects in quantum information framework.

In a previous work \cite{Alexander}, we studied the quantum teleportation in an isotropic expanding spacetime. It was show that the expansion leads to the reduction of the quality of the teleportation, as measured by the fidelity. However, in a realistic situations, it is worth investigating the influence of an anisotropic perturbation of the metric on a quantum teleportation process in an expanding spacetime. This is the main purpose of this Letter which is organized as follows. In Section 2, we present the model of anisotropic spacetime and explicitly compute the Bogolyubov coefficients for a scalar field. In Section 3, we study the teleportation of a qubit between two comoving observers Alice and Bob in an expanding spacetime. In Section 4, we show how the fidelity of a teleported state between Alice and Bob in the distant future, after spacetime expansion saturates, is affected by the anisotropy. Finally, a summary of the main results fo this work is presented in Section 5.


\section{The Model}

In a general curved spacetime, the Lagrangian density for a massive scalar field $\phi(x)$ is given by 
\begin{align} \label{L}
 \mathcal{L} &= \frac{1}{2}\sqrt{-g}[g^{\mu\nu}\partial_{\mu}\phi\partial_{\nu}\phi - (m^2 + \xi R)\phi^2 ] 
\end{align}
where $g^{\mu\nu}$ is the metric with determiant $g$, $R$ is the Ricci scalar curvature and $\xi$ dimensionless paramete. Two values of $\xi$ are of particular interest: for $\xi = 0$, the field is said to be minimally coupled to the metric and for $\xi = \frac{1}{6}$, (\ref{L}) is conformally invariant in the massless limit.
 
 In particular, let us consider a specific model of an anisotropic universe, a Bianchi Type I spacetime, with line element
\begin{align}\label{metric1}
ds^2 = dt^2 - \sum_{j=1}^3 a^2_j(t)dx^2_j,
\end{align} 
  where $a_j(t)$ are arbitrary functions of time. This describe a spatially homogeneous but non-isotropic toy universe. Following \cite{Birrell01, Birrell02, Zeldovich}, we consider the scale factors $a_j(t) = 1 + h_j(t)$ where the perturbation $h_j(t)$ is assumed to be small, i.e., $\mathrm{max}|h_j(t)| \ll 1$. In terms of the conformal time coordinate $d\eta = a^{-1}(t)dt$, the metric (\ref{metric1}) reads
  \begin{align}\label{metric2}
ds^2 = a^2(\eta)\Bigg[d\eta^2 - \sum_{j=1}^3(1 + h_j(\eta))dx_j^2\Bigg]. 
\end{align} 
  As an example, let us now assume $h_j(t)$ to be 
  \begin{align}
  h_j(t) = e^{-\rho\eta^2}\cos(\epsilon\eta^2 + \delta_j),
  \end{align}
  with $\delta_j = \frac{\pi}{2}, \frac{\pi}{2} + \frac{2\pi}{3}, \frac{\pi}{2} + \frac{4\pi}{3}$. It is worthy mention that this choice satisfies the condition $\sum_{j=1}^3h_j(t) = 0$. The dynamics of scalar field $\phi$ in the conformal observer frame  is given by 
 \begin{align} \label{Eq4}
 \partial_{\eta}^2\phi + 2\frac{a'(\eta)}{a(\eta)}\partial_{\eta}\phi - \nabla^2\phi + a^2(\eta)(m^2 + \xi R)\phi = 0,
\end{align}
 where the prime denotes derivatives with respect to conformal time. Because of spatial translation invariance of the spacetime, the solution of the Eq. (\ref{Eq4}) may be written as 
\begin{align}
  \phi_k(x) = (2\pi)^{-\frac{3}{2}}a^{-1}(\eta)e^{i\bold{k}\cdot\bold{x}}f_{k}(\eta),
\end{align}
where to leading order in $h_j$, $f_{k}(\eta)$ satisfies
\begin{align}
\left[\eta^{\mu\nu}\partial_{\mu}\partial_{\nu} + m^2\right]f_{k}(\eta) + V(\eta)f_{k}(\eta) = 0, \label{Eqf}
\end{align}
with 
\begin{align}
V(\eta) &= [a^2(\eta) - a^2(-\infty)]m^2 + (\xi - \frac{1}{6})a^2(\eta)R(\eta) \nonumber \\
&- \sum_{j=1}^3h_j(\eta)k_j^2 . \nonumber
\end{align}
By imposing the conditions \cite{Birrell01, Birrell02}:
\begin{align} \label{Conditions}
\begin{split}
 a^2(\eta) R(\eta) &\to 0  \,\, \text{as} \,\, \eta \to \pm \infty \,\, \text{if} \,\, \xi \ne 1/6,  \\
 h_j(\eta) &\to 0 \,\, \text{as} \,\, \eta \to \pm \infty, \\
 a^2(\eta) &\to a^2(\pm \infty) < \infty \,\, \text{as} \,\, \eta \to \pm \infty \,\, \text{if} \,\, m \ne 0,
 \end{split}
\end{align}
 we may treat $V(\eta)$ as small to solve Eq. (\ref{Eqf}) by iteration to the lowest order in $V(\eta)$ in terms of the momentum space propagator.  Thus the integral form of the Eq. (\ref{Eqf}) becomes
\begin{align}\label{fk}
 f_{k}(\eta) = f^{\mathrm{in}}_{k}(\eta) - \int^{\infty}_{-\infty} G_{r}(\eta,\eta')V(\eta')f_{k}(\eta')d\eta' , 
\end{align}
where $f^{\mathrm{in}}_{k}(\eta)$ is the free-wave solution propagating from the in-region, defined by
$f^{\mathrm{in}}_{k}(\eta) = (2\omega)^{-\frac{1}{2}}e^{-i\omega\eta},$  with $\omega = \sqrt{k^2 + m^2}$. The propagator $G_{r}(\eta,\eta')$ satisfies
\begin{align}
\left[\eta^{\mu\nu}\partial_{\mu}\partial_{\nu} + m^2\right]G_{r}(\eta,\eta') = \delta(\eta - \eta')\, ,
\end{align}
and, in momentum space, reads
\begin{align}
   G_{r}(\eta,\eta') = \frac{1}{2\pi}\int\frac{e^{-ik'_{0}(\eta' - \eta)}}{k'^{2}_{0} - \omega^{2}_{k} - i\varepsilon}dk'_{0}\, . 
\end{align}
 The momentum integral can be performed by closing the integration contour in the upper-half complex momentum plane. However, for the calculation of the Bogoliubov coefficients, it suffices to notice that in the limit $\eta \longrightarrow \infty$, $f_{k}(\eta)$ can be written in terms of the mode functions $f^{\mathrm{in}}_{k}(\eta)$ as
\begin{align}
f^{\mathrm{out}}_{k}(\eta) = (2\omega)^{-\frac{1}{2}}\left[\alpha_{k}e^{-i\omega\eta} + \beta_{k}e^{i\omega\eta}\right], \label{fklater}
\end{align}
 where the Bogoliubov coefficients results as
\begin{align}
\alpha_{k} &= 1 + \frac{i}{\sqrt{2\omega}}\int^{\infty}_{-\infty}e^{i\omega\eta'}V(\eta')f_{k}(\eta')d\eta', \nonumber \\
\beta_{k} &= -\frac{i}{\sqrt{2\omega}}\int^{\infty}_{-\infty}e^{-i\omega\eta'}V(\eta')f_{k}(\eta')d\eta'. \label{Alpbeta}
\end{align}
 To the lowest order in $V(\eta)$ we have $f_{k}(\eta) \cong f^{\mathrm{in}}_{k}(\eta)$, and Bogoliubov coefficients become
\begin{align}
\alpha_{k} &= 1 + \frac{i}{2\omega}\int^{\infty}_{-\infty}V(\eta')d\eta', \nonumber \\
\beta_{k} &= -\frac{i}{2\omega}\int^{\infty}_{-\infty}e^{-2i\omega\eta'}V(\eta')d\eta'. 
\end{align}
Now, let us consider a scale factor $a^2 (\eta )$ which allows us to calculate the Bogoliubov coefficients and satisfy the conditions (\ref{Conditions}):
\begin{align} 
  a^2 (\eta ) = 1 + \epsilon(1 + \tanh(\rho\eta)), \label{metric2}
\end{align}
where the rate of spacetime expansion is given by the parameter $\rho$ and the total amount of expansion by $\epsilon$. $a^2(\eta)$ represents conveniently a spacetime that undergoes a period of smooth expansion and becomes flat in the distant past (in-region)($\eta \rightarrow -\infty$) and in the far future (out-region)($\eta \rightarrow +\infty$). By inserting in $\alpha_k$ and $\beta_{k}$ the explicit form of $V(\eta)$ and $a^2(\eta)$, we can write $\alpha_k$ and $\beta_{k}$ as
\begin{align} \label{betak}
\begin{split}
\alpha_k &= 1 + \alpha_k^{(m)} + \alpha_k^{(\xi)} + \alpha_k^{(h)}, \\
\beta_k &= \beta_{k}^{(m)} + \beta_{k}^{(\xi)} + \beta_{k}^{(h)},
\end{split}
\end{align}
where to leading order in $\epsilon$ we find 
\begin{align} \label{alphakm}
\alpha_k^{(m)} &= \frac{i\epsilon m^2}{\omega}, \\ 
\beta_{k}^{(m)} &= -\frac{m^2\epsilon}{2\omega\rho}\frac{\pi}{\sinh(\frac{\pi\omega}{\rho})}, \label{betakm} \\ 
\alpha_k^{(\xi)} &= (\xi - \frac{1}{6})\frac{i\rho}{12\epsilon\omega}[(\epsilon + 1)\ln(2\epsilon + 1) - 2\epsilon], \label{alphakxi} \\ 
\beta_{k}^{(\xi)} &= (\xi - \frac{1}{6})\frac{\omega\epsilon}{6\rho}\frac{\pi}{\sinh(\frac{\pi\omega}{\rho})}, \label{betakxi} \\ 
\alpha_k^{(h)} &= \frac{i\sqrt{\pi}}{2\omega}\sum_{j=1}^3k_j^2\mathrm{Re}\Bigg(\frac{e^{-i\delta_j }}{\sqrt{\rho + i\epsilon})}\Bigg), \label{alphakh} \\ 
\beta_{k}^{(h)} &= -\frac{\sqrt{\pi}}{2\omega}\sum_{j=1}^3k_j^2\mathrm{Re}\Bigg(\frac{e^{-i\delta_j + i\frac{\pi}{2} -\frac{\omega^2}{\rho + i\epsilon}}}{\sqrt{\rho + i\epsilon})}\Bigg). \label{betakh}
\end{align}
 Notice that the Bogoliubov coefficients have direct dependence on the mass $m$, on the coupling constant $\xi$ and on the anisotropy parameter $h_j$. In the following we will investigate the influence of each of these independent contribution on a quantum teleportation protocol. 


\section{Quantum Teleportation}

Let us consider the following gedanken experiment: two comoving observers Alice and Bob 
in an expanding spacetime shared the maximally entangled in the distant past
\begin{align} \label{past_Bell}
\vert \beta \rangle = \frac{1}{\sqrt{2}}\left(\vert \textbf{0}_{A}^{\mathrm{in}}\rangle\vert\textbf{0}_{B}^{\mathrm{in}}\rangle + \vert \textbf{1}_{A}^{\mathrm{in}}\rangle\vert\textbf{1}_{B}^{\mathrm{in}}\rangle\right).
\end{align}
 where states $\vert\textbf{0}_{A}\rangle$ and $\vert \textbf{1}_{A}\rangle$ are defined in terms of the dual-rail basis as suggested in \cite{Alsing02}, $\vert \textbf{0}_{A}\rangle = \vert 1_{A_{1}}\rangle\vert 0_{A_{2}}\rangle$, $\vert \textbf{1}_{A}\rangle = \vert 0_{A_{1}}\rangle\vert 1_{A_{2}}\rangle$, with similar expression for Bob's cavity. Here we suppose that each observed supports the orthogonal modes $\bold{k}_1 $, $\bold{k}_2 $ of the same frequency of a massive scalar field, labeled respectively as $A_{i}$ and $B_{i}$, $i = 1,2 $.
 
\begin{figure}
	\begin{center}
		\includegraphics[scale=1.0]{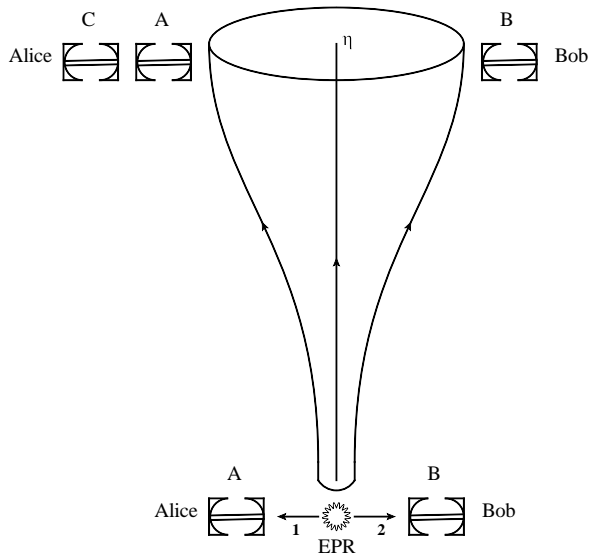}
		\caption{Illustration of a quantum teleportation protocol between two comoving observers in an expanding spacetime. In the distant past, two comoving observers Alice and Bob shared a maximally entangled state. Furthermore, Alice possesses an additional qubit that she wants to teleport to Bob in the distant future, after spacetime expansion saturates.}
	\end{center} \label{fig1}
\end{figure}

As depicted in Fig. 1, Bob perceives Alice receding away from him in his local inertial frame during the period of spacetime expansion. Besides, Bob must be confronted with the fact that this vacuum state becomes an two-mode squeezed state in the distant future:  
\begin{align}
\Ket{0^{\mathrm{out}}_{B_i}} = \sqrt{1 - \gamma_k}\sum_{n=0}^{\infty}\left(\frac{\beta_k}{\alpha_k}\right)^n\vert n_{\bold{k}_i}, n_{-\bold{k}_i}\rangle, \nonumber
\end{align}
 where $\gamma_k = \left|\frac{\beta_k}{\alpha_k}\right|^2$. Similarly, the one-particle state becomes
\begin{align}
\Ket{1_{B_i}^{\mathrm{out}}} = \left(1 - \gamma_k\right)\sum_{n=0}^{\infty}\left(\frac{\beta_k}{\alpha_k}\right)^n\sqrt{n + 1}\vert n + 1_{\bold{k}_i}, n_{-\bold{k}_i}\rangle . \nonumber 
\end{align}
In order to establish the standard teleportation protocol, let us assume a unknown qubit, in Alice's possession, that she wants to teleport to Bob. This qubit state may be written in the dual-rail bases as: $\vert\psi\rangle = a\vert \textbf{0}\rangle + b\vert \textbf{1}\rangle$. In the total system, the full input state is given by $\vert\psi_{0}\rangle = \vert\psi\rangle\vert\beta\rangle$. In the following, Alice will then make a local joint measurement on her two qubits in the distant future. According to the outcome Alice's measurement, the Bob state would be projected into
\begin{align}
\rho^{\bold{k}_1 ,\bold{k}_2 }_{ij} = \Tr_{-\bold{k}_1 ,-\bold{k}_2 }\left(\vert\phi_{i,j}^{\mathrm{out}}\rangle\langle\phi_{i,j}^{\mathrm{out}}\vert\right)
\end{align}
 where $\vert\phi_{i,j}^{\mathrm{out}}\rangle = x_{ij}\vert\textbf{0}_B^{\mathrm{out}}\rangle + y_{ij}\vert\textbf{1}_B^{\mathrm{out}}\rangle$ and there are four possible conditional state amplitudes given by $(x_{00},y_{00}) = (a,b)$, $(x_{01},y_{01}) = (b,a)$, $(x_{10},y_{10}) = (a,-b)$, $(x_{11},y_{11}) = (-b,a)$. The density matrix $\rho^{\bold{k}_1 ,\bold{k}_2 }_{ij}$ has a block diagonal form. Considering the density matrix for the $n = \lbrace 0, 1, 2\rbrace$ excitation sectors, we find

\begin{align*}
  \rho^{\bold{k}_1 ,\bold{k}_2 }_{ij} = \left(1 - \gamma_k\right)^3
  &\left(\begin{matrix}
  0 & 0 & 0 \\
  0 & \rho^{1} & 0 \\
  0 & 0 & \rho^{2} 
  \end{matrix}\right),
\end{align*}
with
\begin{align*}
\rho^{1} = \left(\begin{matrix}
  |x_{ij}|^2 & x_{ij}y_{ij}^* \\
   x_{ij}^*y_{ij}& |y_{ij}|^2 \\
  \end{matrix}\right),
\end{align*}
\begin{align*}
\rho^{2} =  \left(\begin{matrix}
  2\gamma_k|x_{ij}|^2 & \sqrt{2}\gamma_k x_{ij}y_{ij}^* & 0 \\
  \sqrt{2}\gamma_k x_{ij}^*y_{ij} & \gamma_k & \sqrt{2}\gamma_k x_{ij}y_{ij}^* \\
   0 & \sqrt{2}\gamma_k x_{ij}y_{ij}^* & 2\gamma_k|y_{ij}|^2
  \end{matrix}\right).
\end{align*}
 Here we have used the basis $\lbrace |00\rangle, |01\rangle, |10\rangle, |02\rangle, |11\rangle, |20\rangle\rbrace$. Notice that the excitation sectors has probability: $p_{0} = 0$, $p_{1} = (1 - \gamma_k)^3$, and $p_{2} = (1 - \gamma_k)^3\gamma_k$. Finally, the teleportation will be complete when Alice send the result of her measurement $\left\lbrace i,j\right\rbrace$ to Bob through a classical channel.  After Bob receives the message from Alice, he performs a unitary operation on his qubit to transform it to the desired state $\vert\psi\rangle$. It is noteworthy to mention that if Alice makes the measurement in the distant past but Bob waits until the distant future before receiving Alice’s message and applying the relevant unitary, then she does not need to remain in causal contact; in this case, all is needed is that he gets the measurement’s result before falling out of causal contact. The results one would obtain for the fidelity under this setup are however the same. In addition, the roles of Alice and Bob may be exchanged, as they are both comoving observers.
 
 The performance of the teleportation is quantified by the teleported state fidelity $F_k = \Tr\left(\vert\psi\rangle\langle\psi\vert\rho^{\bold{k}_1 ,\bold{k}_2 }_{ij}\right)$, defined as the overlap of the teleported state $\vert\psi\rangle$ and the density matrix $\rho^{\bold{k}_1 ,\bold{k}_2 }_{ij}$. In terms of Bogoliubov coefficients $\beta_k$ and $\alpha_k$, the fidelity turns out to be
 \begin{align}\label{fidelity}
   F_k(\rho, \epsilon, h) &= \left( 1 - \left|\frac{\beta_k}{\alpha_k}\right|^2\right)^3, \nonumber \\
   & = \left(1 - \left|\frac{\beta_{k}^{(m)} + \beta_{k}^{(\xi)} + \beta_{k}^{(h)}}{1 + \alpha_k^{(m)} + \alpha_k^{(\xi)} + \alpha_k^{(h)}}\right|^2\right)^3
\end{align}
 This result, together with (\ref{alphakm}, \ref{betakm}, \ref{alphakxi}, \ref{betakxi}, \ref{alphakh}, \ref{betakh}), allows us to determine the influence of the anisotropic perturbation on a teleportation protocol in an expanding spacetime.
 

\section{Results and Discussion}

 In order to realize the plots, let us assume that wavevector $\overrightarrow{k} = (k_1, k_2, k_3)$ has a general direction specified by spherical coordinates $(k, \theta, \phi)$ as
 \begin{align*}
 \overrightarrow{k} = k\sin\theta\cos\phi \hat{x} + k\sin\theta\sin\phi \hat{y} + \cos\theta \hat{z},
 \end{align*}
 where $k^2 = k_1^2 + k_2^2 + k_3^2$. Notice that the effects of the anisotropy is expected to depend upon the direction of the particle momentum. Thus, the influence of anisotropy on the fidelity of a teleported state between Alice and Bob in the distant future can be quantified by the azimuthal angle $\theta$ and/or the polar angle $\phi$.

\begin{figure}
	\centering
	\includegraphics[height=4cm]{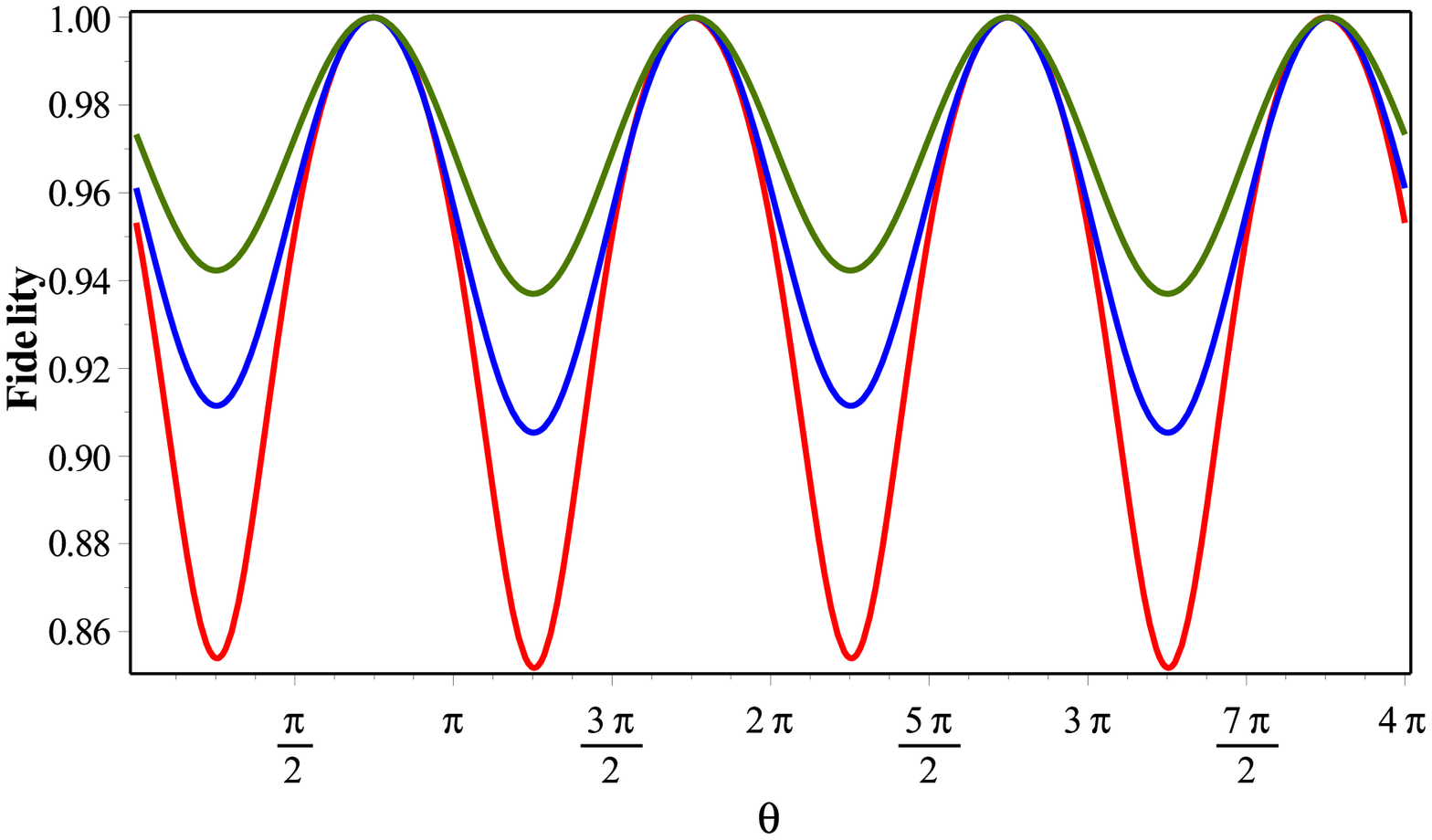} \\ (a) \\  \includegraphics[height=4cm]{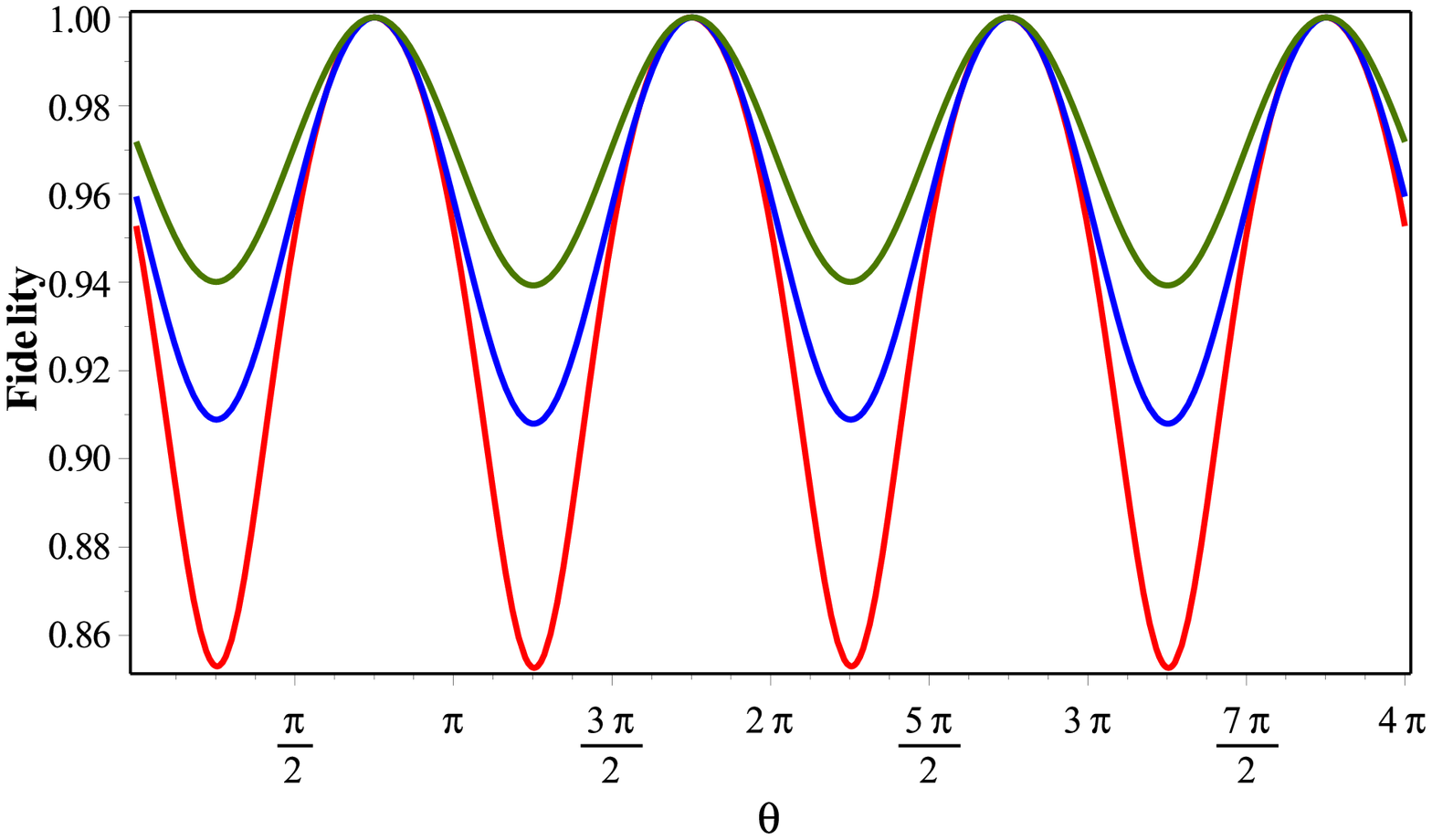} \\ (b)
	\caption{Fidelity as a function of the azimuthal angle $\theta$ for different values of the parameter $\rho$: $\rho = 1$ (red line), $\rho = 5$ (blue line) and $\rho = 10$ (green line). Here we have fixed $k = 1$, $m = \epsilon = 0.1$ and $\phi = \frac{\pi}{2}$. The plots (a) and (b) show the results for the minimal coupling ($\xi = 0$) and conformal coupling ($\xi = \frac{1}{6}$).}
	\label{fig2}
\end{figure}

The fidelity (\ref{fidelity}) have been plotted in Fig. (\ref{fig2}) as a function of azimuthal angle $\theta$, for different values of parameter $\rho$. We can observer from Fig. (\ref{fig2}) that the spectral behavior of the fidelity oscillates as a function of $\theta$. This shows that the fidelity is sensible to the direction of the particle momentum. 

 Other interesting thing to note in Fig. (\ref{fig2}) is that the amplitude of the oscillation due to anisotropy decrease with the cosmological parameter $\rho$ grows. This suggests, as expected, that the anisotropic effects on the fidelity becomes more significative in the regime of smooth expansion.

\begin{figure}
	\centering
	\includegraphics[height=4cm]{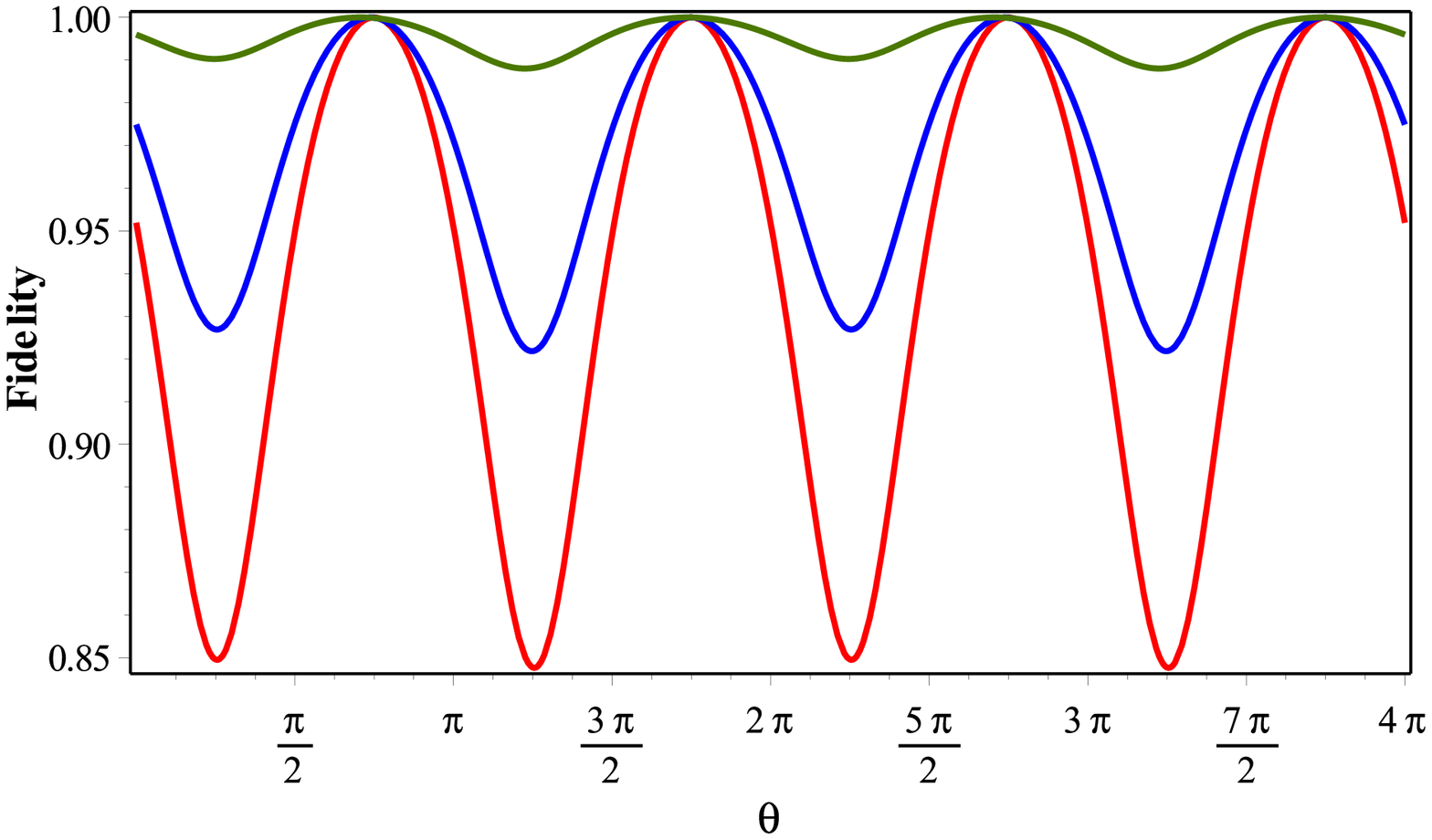} \\ (a) \\ \includegraphics[height=4cm]{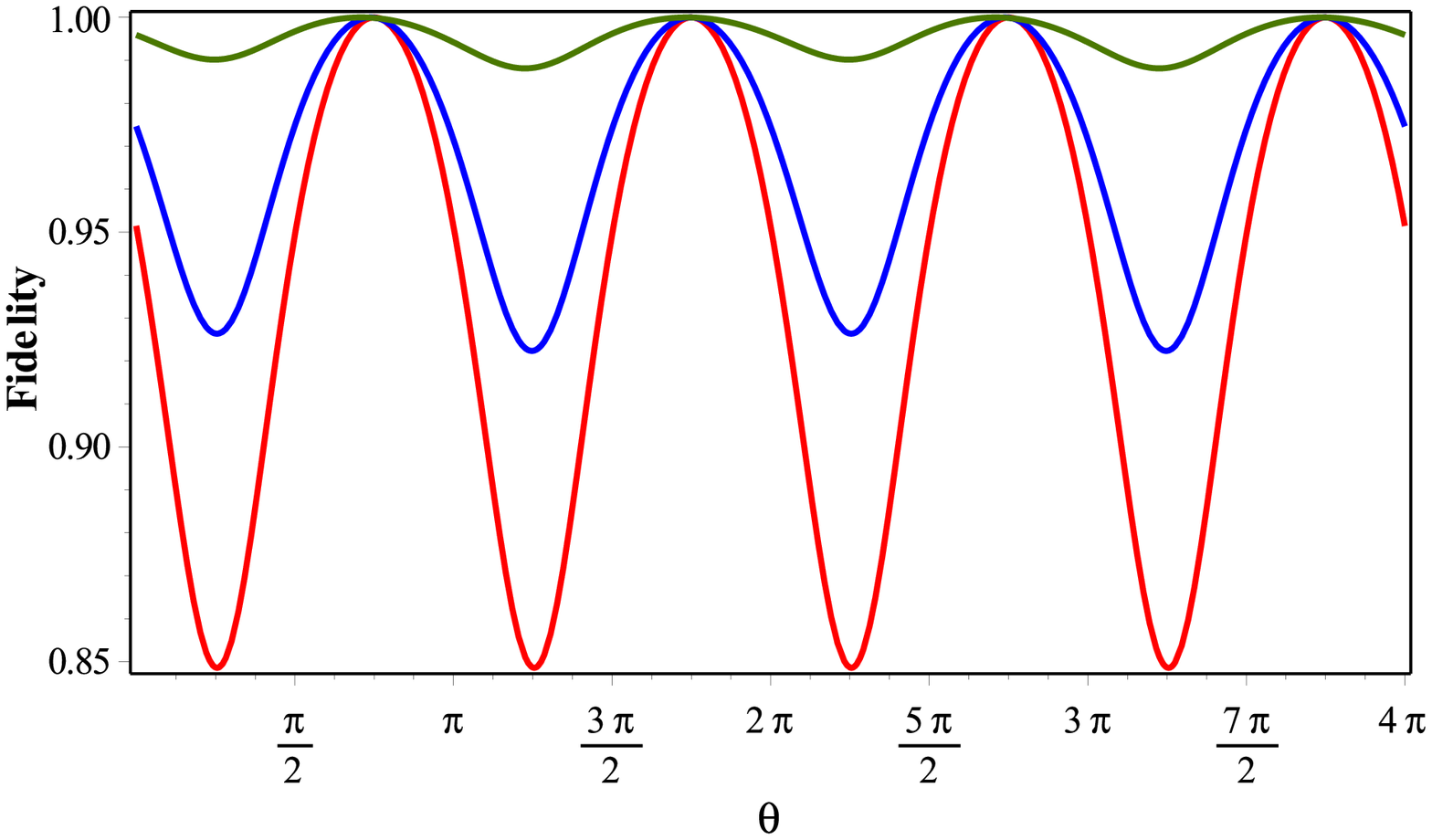} \\ (b)
	\caption{Fidelity as a function of the azimuthal angle $\theta$ for different values of the parameter $m$: $m = 0$ (red line), $m = 0.5$ (blue line) and $m = 1$ (green line). Here we have fixed $k = 1$, $\rho = 1$, $\epsilon = 0.1$ and $\phi = \frac{\pi}{2}$. The plots (a) and (b) show the results for the minimal coupling ($\xi = 0$) and conformal coupling ($\xi = \frac{1}{6}$).}
	\label{fig3}
\end{figure}

 Let us now examine the influence of mass on the performance of our protocol. In Fig. (\ref{fig3}) we can see that the amplitude of the oscillation of fidelity decrease as $m$ increases. Thus, the contribution of the anisotropic perturbation is expected to be more significative when particles are light. 

\begin{figure}
	\centering
	\includegraphics[height=4cm]{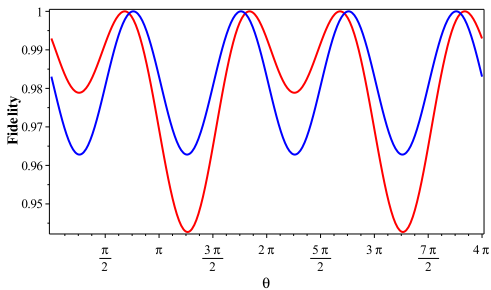}
	\caption{Fidelity as a function of the azimuthal angle $\theta$ for different values of the parameter $\xi$: $\xi = 0$ (red line) and $\xi = 1/6$ (blue line). Here we have fixed $m = 0$, $k = 1$, $\rho = 20$, $\epsilon = 1$ and $\phi = \frac{\pi}{2}$.}
	\label{fig4}
\end{figure}

  Concerning the role played by the parameter of the coupling $\xi$ on the fidelity. Figures 2 and 3 show that the influence of curvature coupling becomes imperceptible in the regime of smooth expansion, i.e., when $\frac{\rho}{\omega} \leq 1$. In particular, particles with mass $m = 1$ and momentum $k = 1$, the regime of smooth expansion correspond to values of $\rho \leq 1.41$. Only in the regime of fast expansion (values of $\frac{\rho}{\omega} \gg 1$) which implies in the limit of massless particles and small values of momentum $k$, the contribution of coupling becomes more relevant (or noticeable) on the spectral behaviour of fidelity, see Fig. (\ref{fig4}).

 The above results suggest that there is a privileged value of azimuthal angle $\theta$ for which the expansion of spacetime leads to a reduction of the fidelity. Notice that for the polar angle $\phi = \frac{\pi}{2}$ and the azimuthal angle $\theta \approx \frac{3\pi}{4} + n\pi$ with $n = 0,1,2,...$ the qubit of Alice can be teleported with maximum value of fidelity $F \approx 1$. On the other hand, if the azimuthal angle is chosen such that $\theta \neq \frac{3\pi}{4} + n\pi$, then the efficiency of the process decreases. We benefit from this special behavior that the fidelity presents at a specific direction of the momentum to design method to extract information on possible consequences of the presence of anisotropic perturbation on quantum communication in an expanding spacetime.

\section{Conclusions} 

In summary, we have studied a quantum teleportation process between two comoving observers in an anisotropic expanding spacetime. An oscillation of fidelity in function of the azimuthal angle has been observed. We found that for the polar angle $\phi = \frac{\pi}{2}$ and the azimuthal angle $\theta \neq \frac{3\pi}{4} + n\pi$ with $n = 0, 1, 2, ...$ the efficiency of the process decreases, i.e., the fidelity is less than one. In addition, we show that the anisotropic effects on the fidelity becomes more significative in the regime of smooth expansion and the limit of massless particles. On the other hand, we found that the influence of curvature coupling becomes noticeable in the regime of fast expansion (values of $\frac{\rho}{\omega} \gg 1$). Our results suggest that the presence of a small disturbance in the metric (anisotropic perturbation) affects significantly the performance of a teleportation protocol in an expanding spacetime.

\section*{Acknowledgments}

HASC would like to thank the Brazilian funding agency CAPES for financial support. PRSC would like
to thank CNPq (Brazilian funding agency) through grant Universal-431727/2018.

\end{document}